\documentclass[a4paper,11pt]{article}
\pdfoutput=1 

\usepackage{jcappub} 

\usepackage[T1]{fontenc} 

\title{Primordial black holes as a dark matter candidate in theories with supersymmetry and inflation}

\author[a,1]{Marcos M. Flores,\note{Corresponding author.}}
\author[a,b]{Alexander Kusenko,}

\affiliation[a]{Department of Physics and Astronomy, University of California, Los Angeles \\ Los Angeles, California, 90095-1547, USA}
\affiliation[b]{Kavli Institute for the Physics and Mathematics of the Universe (WPI), UTIAS \\The University of Tokyo, Kashiwa, Chiba 277-8583, Japan}

\emailAdd{mmflores@physics.ucla.edu}
\emailAdd{kusenko@g.ucla.edu}

\abstract{
We show that supersymmetry and inflation, in a broad class of models, generically lead to formation of primordial black holes (PBHs) that can account for dark matter. Supersymmetry predicts a number of scalar fields that develop a coherent condensate along the flat directions of the potential at the end of inflation.  The subsequent evolution of the condensate involves perturbative decay, as well as fragmentation into Q-balls, which can interact by some  long-range forces mediated by the scalar fields.  The attractive scalar long-range interactions between Q-balls  facilitates the growth of Q-balls until their ultimate collapse to black holes.  For a flat direction lifted by supersymmetry breaking at the scale $\Lambda\sim 100\, {\rm TeV}$, the black hole masses are of the order of $(M_{\rm Planck}^3/\Lambda^2)\sim 10^{22}\, {\rm g}$, in the allowed range for dark matter.  Similar potentials with a lower scale $\Lambda$ (not necessarily associated with supersymmetry) can result in a population of primordial black holes with larger masses, which can explain some recently reported microlensing events. 
}

\begin{document}
\maketitle
\flushbottom

\section{Introduction}
\label{sec:intro}

Astrophysical black holes are known to have formed in recent cosmological times. However, black holes could have also formed in the early Universe. These primordial black holes (PBHs) can account for all or part of dark matter~\cite{Zeldovich:1967,Hawking:1971ei,Carr:1974nx,Khlopov:1985jw,Dolgov:1992pu,Yokoyama:1995ex,Wright:1995bi,GarciaBellido:1996qt,Kawasaki:1997ju,Green:2004wb,Khlopov:2008qy,Carr:2009jm,Frampton:2010sw,Kawasaki:2016pql,Carr:2016drx,Inomata:2016rbd,Pi:2017gih,Inomata:2017okj,Garcia-Bellido:2017aan,Georg:2017mqk,Inomata:2017vxo,Kocsis:2017yty,Ando:2017veq,Cotner:2016cvr,Cotner:2017tir,Cotner:2018vug,Sasaki:2018dmp,Carr:2018rid,Germani:2018jgr,Banik:2018tyb,Escriva:2019phb,Germani:2019zez,1939PCPS...35..405H,Cotner:2019ykd,Kusenko:2020pcg,deFreitasPacheco:2020wdg,Takhistov:2020vxs,Biagetti:2021eep}. In addition, PBHs can be responsible for a variety of astrophysical phenomena including the gravitational events detected at LIGO~\cite{Abbott:2016blz,Abbott:2016nmj,Abbott:2017vtc,Clesse:2016vqa,Bird:2016dcv,Sasaki:2016jop},
the formation of supermassive black holes \cite{Bean:2002kx,Kawasaki:2012kn,Clesse:2015wea} or $r$-process nucleosynthesis \cite{Fuller:2017uyd}.
Some models of PBH formation rely on supersymmeric flat directions~\cite{Cotner:2016cvr,Cotner:2017tir,Cotner:2019ykd}.
Scalar fields associated with supersymmetry can develop a large value at the end of inflation, and then  fragment~\cite{Kusenko:1997si} into lumps of scalar fields, Q-balls, which can become the building blocks of PBHs through their gravitational interactions~\cite{Cotner:2016cvr,Cotner:2016dhw,Cotner:2019ykd}.  

We will describe a much more efficient mechanism for PBH formation. In addition to gravitational interactions, the Q-balls can have much stronger scalar interactions which mediate relatively long-range forces.  These attractive interactions can lead to PBH formation similar to the Yukawa interactions between fermions~\cite{Flores:2020rc}.  The two main effects of the long-range scalar forces are the halo formation (possible even in the radiation dominated era~\cite{Amendola:2017xhl,Savastano:2019zpr,Flores:2020rc,Domenech:2021uyx}) and the radiative cooling by emission of scalar waves~\cite{Flores:2020rc,Flores:2021sp}.  
We will show that such interactions lead to PBH formation in a broad class of supersymmetric models.  Furthermore, the black hole mass is related to supersymmetry breaking scale and is naturally in the range where PBHs are allowed to account for 100\% of dark matter.  The experimental limits on a ``fifth force" in the relevant range are very weak.  The current experimental bounds allow a fifth force $10^{12}-10^{18}$ times the strength of the force of gravity on the length scales $(10^{-8} - 10^{-6})$~m relevant for PBH formation~\cite{Adelberger:2003zx,Heacock:2021btd}.

Formation of a supersymmetric scalar condensate after inflation and its subsequent fragmentation is a well understood phenomenon~\cite{Kusenko:1997si}.  The scalar potential corresponding to a superpotential $W$ can be written as 
\begin{equation}
U_{\rm SUSY}=\sum_i |F_i|^2+ \sum_a \frac{g^2_a}{2} |D_a|^2, 
\end{equation}
where $F_i=\partial W/\partial \phi_i$, $D_a=\phi_i^\dagger T_a^{ij}\phi_j$, and the sum runs over all the chiral superfields $\phi_i$ and all the gauge generators $T_a$.  This potential has flat directions for the linear combinations of fields that make both the $F$-terms and the $D$-terms vanish.  D-flatness implies that the flat direction is  parameterized by a  gauge-invariant combination of the scalar fields, and F-flatness imposes an additional constraint, which still allows a large number of independent solutions in  the Minimal Supersymmetric Standard Model (MSSM)~\cite{Gherghetta:1995dv} or other generalization of the Standard Model.  After the supersymmetry breaking is included, the flat directions are lifted by the soft supersymmetry breaking terms in $U_{\rm soft}$: 
\begin{equation}
U(\phi)=U_{\rm SUSY}+U_{\rm soft}
\end{equation}
Soft supersymmetry breaking does not contain any quartic terms, so the flat direction parameterized by some field $\varphi $ remains a slowly growing direction in the potential.  In the early Universe, this implies that it develops a large VEV.  There are two reasons for this: (i) the effective minimum of the potential in de Sitter space can be displaced because of the terms $\propto H^2 \varphi^2$ that come from the K\"ahler potential~\cite{Affleck:1984fy,Dine:1995kz} and (ii) regardless of where the minimum is, the field undergoes fluctuations away from the minimum~\cite{Chernikov:1968zm,Tagirov:1972vv,Bunch:1978yq,Linde:1982uu,Lee:1987qc,Starobinsky:1994bd}. The consequence of (i) is that, if the mass $m_\varphi $ is small and the effective  contribution from the K\"ahler potential is {\it negative}, $\sim ( -|{\rm const}|) H_{\rm infl}^2 \varphi^2 $, the scalar field's effective potential can have a minimum at some large value.  However, the fluctuations (ii) can also lead to a large VEV for a flat direction.   The results of Refs.~\cite{Chernikov:1968zm,Tagirov:1972vv,Bunch:1978yq,Linde:1982uu,Lee:1987qc,Starobinsky:1994bd} can be summarized as follows: each scalar degree of freedom, on average, carries the amount of energy density of the order of $H_{\rm infl}^4$, where $H_{\rm infl}$ is the Hubble parameter during inflation. (Thus one can think of $H_{\rm infl}$ as the effective temperature of de Sitter space for the purposes of scalar fluctuations.) Since the flat directions are lifted only slightly by the supersymmetry breaking terms, the corresponding VEVs have to be large to provide them with the energy density $\sim H_{\rm infl}^4$. After inflation is over, and the Hubble parameter decreases, the scalar condensate is subject to instabilities that lead to the formation of SUSY Q-balls~\cite{Kusenko:1997si,Dine:2003ax,Enqvist:2003gh}.
The gas of Q-balls behaves as matter with relatively few particles per volume, which implies relatively large fluctuations capable of creating PBH with the masses and abundances suitable for dark matter~\cite{Cotner:2016cvr,Cotner:2016cvr,Cotner:2016dhw,Cotner:2017tir,Cotner:2018vug,Cotner:2019ykd}.  This channel for PBH formation is based on gravitational interaction.  Some other possibilities exist for PBH formation from Affleck--Dine baryogenesis~\cite{Dolgov:2008wu,Hasegawa:2017jtk,Kawasaki:2021zir}.

\section{PBHs from long-range scalar forces and scalar cooling}

\subsection{SUSY Q-balls and scalar interactions}

Let us now describe a new PBH formation mechanism which is even more efficient.  Q-balls generically couple to scalar fields mediating long-range forces that can be stronger than gravity.  For example, a Q-ball formed along an $LLe$ flat direction (in the notation of Ref.~\cite{Gherghetta:1995dv}) is built from the scalar fields that couple to the Higgs boson.  Outside the Q-ball the Higgs boson can have a mass that is small enough to allow attraction between Q-balls on the length scales smaller than the Higgs Compton wavelength. In addition, some independent flat direction, parameterized by a scalar field of small mass (by virtue of the flatness of the potential), can couple to a given flat direction by the effective couplings induced through loops.  
Such scalar fields can mediate attractive scalar interactions between Q-balls.  
	
To avoid complications with the multiple fields of MSSM, we will focus on a simple model that captures the relevant physics of SUSY Q-ball interactions.  We assume that two flat directions, $\phi$ and $\chi$ ($U_{\rm SUSY}(\phi)=U_{\rm SUSY}(\chi)=0$)  are lifted by some soft terms that arise from a  gauge-mediated supersymmetry breaking~\cite{deGouvea:1997afu,Dvali:1997qv,Kusenko:1997si,Hong:2017qvx,Kawasaki:2019ywz,Kasuya:2000sc}: 

\begin{equation}
U_{\rm soft}(\phi) \approx \Lambda^4_{\rm SUSY} 
\left (\log\left ( 1+\frac{|\phi|^2}{M^2_{\rm mess}}\right )
\right )^2+V_{\rm grav}+...
\end{equation}
\begin{equation}
U_{\rm soft}(\chi) \approx \Lambda^4_{\rm SUSY} 
\left (\log\left ( 1+\frac{|\chi|^2}{M^2_{\rm mess}}\right )
\right )^2+V_{\rm grav}+...
\end{equation}
We will assume that the gauge mediated terms dominate over the gravity mediated terms, so that $V_{\rm grav}$ can be neglected. 

Higher dimensional operators can induce an interaction term that couples the two flat directions: 
\begin{equation}
V_{\chi\phi}(\phi,\chi)
=
-y\chi\phi^\dagger\phi+ {\rm h.c.} 
\label{eqn:crossterm}
\end{equation}
where the $y$ is an effective (dimensionful) coupling. 

We also assume that the field $\chi $ has zero charge with respect to the U(1) symmetry responsible for the stability of the Q-balls made of the $\phi$ fields.   It is clear that the coupling in Eq.~(\ref{eqn:crossterm}) preserves the U(1) symmetry 
$$
\phi \rightarrow e^{i \theta} \phi, \ \ \chi \rightarrow \chi
$$
Due to the difference in the equations of motion and the initial conditions (induced by the higher-dimension operators, as well as CP violating A-terms~\cite{Dine:1995kz}), the fields $\phi$ and $\chi$ can, in general have very different ``angular velocities".   It is well-known that a flat direction can remain homogeneous in space unless the global charge density exceeds some critical value. For a larger charge density, fragmentation into Q-balls takes place~\cite{Kusenko:1997si}.  It is possible, therefore, for the $\phi$ field fragment into Q-balls, while the $\chi$ field remains spatially homogeneous.

Specifically, we assume that
\begin{eqnarray}
\phi &=& |\bar \phi(x,t) | \exp \{i \omega_\phi t \} \\
\chi &=& |\bar \chi (x,t) | \exp \{i \omega_\chi t \},
\end{eqnarray}
and we consider the range of parameters for which $\omega_\chi =0$ (for example, if the interaction term in Eq.(\ref{eqn:crossterm}) precludes any rotation or $\chi$, while it always preserves the U(1) symmetry of the $\phi $ field), or $\omega_\chi < |U''| < \omega_\phi $ during the relevant part of the evolution of the scalar condensate.  As long as this is the case, the two flat directions behave differently.  There is a band of unstable modes in the $\phi$ field with wavenumbers $k$ in the range $H<k<\sqrt{\omega^2_\phi-U''}$, and the $\phi$ condensate evolves into Q-balls~\cite{Kusenko:1997si}.  In contrast, the field $\chi$ is not subject to an instability because $\omega^2_\phi-U'' <0 $, and, therefore,
\begin{eqnarray}
\partial_x \bar \chi=0 \\
\partial_x \bar \phi \neq 0 .
\end{eqnarray}
Fragmentation of the $\phi$ field leads to Q-balls with a global charge $Q$ and 
\begin{eqnarray}
\bar \phi\sim \left  \{ 
\begin{array}{ll}
     0, & {\rm outside\ a\ Q-ball}  \\
      \Lambda_{\rm SUSY} Q^{1/4}, &
     {\rm inside\ a\ Q-ball} 
\end{array} 
\right . 
\end{eqnarray}
In contrast, 
\begin{eqnarray}
\bar \chi\sim  \chi_0, \  {\rm inside \ and\   outside\ a\ Q-ball} 
\end{eqnarray}
The effective mass squared of the $\chi$ field is 
\begin{eqnarray}
m^2_\chi\sim  \partial^2 U_{\rm soft}(\chi)/\partial \chi^2+c H^2+... \\
\sim  - \Lambda^4_{\rm SUSY}/\chi_0^2 +c H^2
\sim c H^2
, 
\end{eqnarray}
where the contribution of the scalar potential is very small and negative due to the ``tachionic" shape of the potential, while the  K\"ahler potential terms $\sim c H^2$ ($|c| \sim 1$) can be important.  It is reasonable to assume  $m_\chi \sim H$, which implies that the corresponding Compton wavelength is large enough for the interaction Eq.(\ref{eqn:crossterm}) to mediate a long-range force on subhorizon scales.

The properties of Q-balls in our scenario can be described in terms of the global charge $Q$:
\begin{equation}
\label{eq:Qrelations}
M_Q \sim \Lambda |Q|^{\alpha},
\qquad
R_Q \sim \frac{|Q|^{\beta}}{\Lambda},
\qquad
\omega_Q \sim \Lambda Q^{\alpha - 1}.
\end{equation}
where $M_Q$ and $R_Q$ are the mass and radius of a soliton with charge $Q$ and $\omega_Q$ is the energy per charge.

The cubic interaction with a dimensionful coupling $y$ can arise from supersymmetry preserving terms (such as the $\mu$ term), or supersymmetry breaking terms (such as the $A$ term)~\cite{Dine:2003ax}.  In the latter case, the loop corrections for the mass of the $\chi$ field can be important.  The parameters $\alpha$ and $\beta$ are positive and less than or equal to one, and $\Lambda$ is the energy scale associated with the scalar potential. For flat direction lifted by gauge-mediated supersymmetry breaking,  $\alpha = 3/4$ and $\beta = 1/4$~\cite{Dvali:1997qv,Kusenko:1997si}.  Other types of supersymmetry breaking lead to different values of $\alpha$ and $\beta$~\cite{Enqvist:2003gh}. 
While the field $\chi$ can receive an effective mass $m_\chi \sim H$, of the order of the Hubble parameter~\cite{Kawasaki:2011zi}, on the length scales smaller than $m_\chi^{-1} \sim H^{-1} $, it acts as a long-range attractive force. 

\subsection{Halo formation and scalar cooling channels}

The system of interacting Q-balls is similar to the system of fermions interacting via Yukawa forces~\cite{Amendola:2017xhl,Savastano:2019zpr,Flores:2020rc,Domenech:2021uyx,Flores:2021sp}.  This system can form halos during the radiation or matter dominate era, and the halos can collapse into the black holes if the radiative cooling by emission of $\chi$ waves is efficient enough~\cite{Flores:2020rc}.  We introduce a dimensionless coupling $g = y/\omega_Q$, which determines the interaction strength.  
A virialized halo of $N$ Q-balls has total energy $E\sim g^2Q_{\rm tot}^2/R$ where $R$ is the characteristic radius. This energy needs to be removed from the halo before a black hole can form.  The system of solitons can radiate $\chi$ waves in several different ways.
	
First, scattering of individual solitons could generate scalar bremsstrahlung. This processes is similar to free-free emission of photons from plasma~\cite{Maxon:1967,Maxon:1972} and is particularly relevant for high-density halos.
	
Second, as the halo contracts it becomes opaque to scalar radiation. When radiation is trapped, cooling continues from the surface. Within the halo, energy transport can occur either through convection or diffusion. The strength of the scalar interactions can easily overpower the viscosity and lead to large Rayleigh numbers. The quick convective timescales aid the surface cooling and facilitate the collapse. 
	
Third, the motion may be incoherent. The power radiated will be proportional to the square of the orbital acceleration $a = \omega^2 R$ where $\omega$ varies for different solitons. This processes is particular applicable to low density configurations.
	
Lastly, the motion may be coherent. In this case, a dipole rotating with a frequency $\omega$ can produce dipole radiation $P_{\rm coh}\propto g^2Q_{\rm tot}^2$. There are various circumstances where coherent radiation may become unimportant. For a system of $N$ identical charges, the center of charge and center of mass coincide implying the dipole moment vanishes. Alternatively, if the mass is proportional to charge (i.e. $\alpha = 1$) then the dipole moment will again vanish. We will treat both of these circumstances as limiting cases for our general framework.
	
To determine the power radiated by coherent motion, we define the ``effective dipole radius" ${\bf r}_Q$ as the difference between the center of charge and center of mass. The rate of energy loss due to coherent motion is 
\begin{equation}
P_{\rm coh} = g^2Q_{\rm tot}^2 \langle |{\bf r}_Q|^2 \rangle\omega^4.
\end{equation}
We will assume charges are specified by the distribution $f_Q(Q)$ with known mean $\langle{Q\rangle}$ and variance $\sigma_Q^2$. From the mass-charge relation, we can also deduce the distribution of masses
\begin{equation}
\label{eq:massdist}
f_M(M)
=
\frac{M^{\frac{1 - \alpha}{\alpha}}}{\alpha\Lambda^{1/\alpha}}
\left[
f_Q((M/\Lambda)^{1/\alpha}) + f_Q(-(M/\Lambda)^{1/\alpha})
\right]
.
\end{equation}
With this distribution, one can calculate $\langle M\rangle$ and $\sigma_M^2$. In addition, we will assume that the spatial distribution of the charges is known. In particular, we will take each spatial coordinate from the distribution $f_r(r)$ with known mean $\langle r\rangle$ and variance $\sigma_r^2$. Without loss of generality, one can assume that $\langle r \rangle = 0$ and that the characteristic length scale for coherent motion is $\sigma_r = R$. Then
\begin{equation}
\label{eq:effdiprad}
\langle|{\bf r}_Q|^2\rangle
\simeq
\frac{3R^2}{N}
\left\{
2 
+
\frac{\sigma_Q^2}{\langle Q\rangle^2} 
+ 
\frac{\sigma_M^2}{\langle M\rangle^2}
-
\frac{2\Lambda\langle Q^{\alpha + 1}\rangle }{\langle M \rangle \langle Q\rangle }
\right\}
\end{equation}
where we have assumed that $N\gtrsim 10$. As expected, this expression vanishes when either $\alpha = 1$ or $f_Q(Q) = \delta(Q - Q_0)$. Altogether, the power radiated for each mode is given by
\begin{align}
P_{\rm brem} 
&\sim 
\frac{16\pi g^8\langle Q \rangle^8 N^3}{M_Q^2R^4}\ln\left(\frac{N \langle Q \rangle^2 g^2}{M_QR}\right)\\[0.25cm]
P_{\rm surf} 
&\sim 
4\pi\frac{g^2\langle Q \rangle^2 N^2}{R^2}\\[0.25cm]
P_{\rm incoh}
&\sim
\frac{g^{10} \langle Q \rangle^{10} N^5}{M_Q^4R^6}\\[0.25cm]
P_{\rm coh} 
&\sim
\frac{g^{10} \langle Q \rangle^{10} N^5}{M_Q^4R^6}\kappa
\end{align}
where $\kappa$ is the bracketed statistical factor in \eqref{eq:effdiprad}. The characteristic time scale for energy loss is defined as
\begin{equation}
\tau = \frac{E}{dE/dt} = \frac{E}{P_{\rm ff} + P_{\rm surf} + \cdots}
.
\end{equation}
When the time scale for energy loss is small compared to the Hubble time at fragmentation, radiative losses are significant. For sufficiently strong scalar forces, all of the solitons enclosed in the Hubble radius collapse into a single charge after a series of mergers. Although mergers decrease the number density, they also increase the average size of the charges. For all of the cooling mechanisms mentioned, mergers will hasten collapse. Mergers begin when the halo radius reaches
\begin{equation}
R_c = N^{1/3}R_Q
\end{equation}
ending in a single $Q$-ball with net charge $N\langle Q\rangle$.

As the horizon increases the newly formed, larger charges  begin to interact and merge just as before. These interaction repeatedly result in formation of a single Q-ball  per horizon.  If the potential remains flat indefinitely, the VEV inside the growing Q-ball can grow greater than the Planck scale, which is not consistent with the field-theoretical description of the Q-balls we employed.  However, the higher-dimensional operators suppressed by the Planck scale can introduce two important effects.  First, the operators induced by gravity need not respect the global symmetries such as the baryon or lepton number conservation.  The Q-ball evolution and growth is then affected by the decay of the Q-ball charge~\cite{Kawasaki:2005xc}.  Second, these operators can lift the flat direction further and prevent the VEV from growing, signaling a transition from a flat-direction Q-ball to a Coleman's original Q-ball with a Q-independent VEV. We, therefore, consider the following set of higher-dimension operators~\cite{Kusenko:2005du}: 
\begin{equation}
V^n(\phi)_{\rm lifting}
\approx
\lambda_n M^4	
\left(
\frac{\phi}{M}
\right)^{n - 1 + m}
\left(
\frac{\phi^*}{M}
\right)^{n - 1 - m}
\end{equation}
These operators lift the flat direction and alter the behavior of the VEV when the charge reaches some value $Q_c$.  For $m\neq 0$, these operators also cause the decay of the global charge inside the Q-ball.  Here, $\Lambda$ is the flat direction height and $M$ is the scale of new physics i.e. $M_{\rm GUT}$ or $M_{\rm Pl}$, etc. and $\lambda_n\sim\mathcal{O}(1)$.  At the critical charge $Q_c$, the properties of the Q ball change to those of curved direction (Coleman) Q balls with $\alpha = 1$ and $\beta = 1/3$. Explicitly,
\begin{equation}
Q_c
\simeq
\lambda_n^{-\frac{2}{n - 1}}
\left(
\frac{M}{\Lambda}
\right)^{\frac{4n - 12}{n - 1}}.
\end{equation}
For $\Lambda\sim 10^3$ GeV and $M\sim 10^3$ GeV, $Q_c\sim 10^{17}$ for $n = 4$ or as large as $10^{32}$ for $n = 7$.

After this point, the charges continue to merge until they reach a maximum charge determined by the equality $R_{s} = 2GM_Q$,
\begin{equation}
Q_{\max}
=
\frac{2^{3/4}\sqrt{3}}{\sqrt{\pi}}Q_c^{1/4}
\left(
\frac{M_{\rm Pl}}{\Lambda}
\right)^3
.
\end{equation}
Once mergers result in charges equal to $Q_{\max}$, the Q-ball becomes a PBH. The newly formed black holes no longer participate in scalar interactions in accordance to the no-hair theorems. 

Since solitons are extended objects, we must constrain the initial the parameters to ensure that PBHs can form. At the time of fragmentation $t_f$, we will assume that $N$ charges form each with identical charge $Q_0$. The constraints discussed below will only apply to the initial population of charges and the fragmentation time.

First, we require that the potential energy density not exceed the critical density. This constrains the maximum number of charges,
\begin{equation}
N \leq \sqrt{4\pi}
\frac{ M_{\rm pl}}{g H_f Q_{0}}
.
\end{equation}
where $H_f$ is the Hubble parameter at $t_f$. 
Second, we require that the initial individual solitons are not massive enough to form BHs, {\it i.e.},  the radius of an individual solition is larger than its Schwarzschild radius: $Q_0 < Q_{\max}$. Third, we will constrain the number of solitons formed after fragmentation by requiring that the energy density of the solitons is less than the total energy. This implies
\begin{equation}
\label{eq:recnstr}
N < \frac{4\pi M_{\rm Pl}^2H_f^{-1}}{\Lambda Q_0^{\alpha}}
.
\end{equation}
\subsection{PBH masses and abundance}

The masses of the PBHs resulting from scalar-cooling of soltions is simply~\cite{Kusenko:2005du},
\begin{equation}
\begin{split}
\label{eq:MPBHEq}
M_{\rm PBH}
& = 
\omega_c Q_{\max}
=
\left(
\pi\sqrt{2}\Lambda Q_c^{-1/4}
\right)Q_{\max}
\\[0.25cm]
&\sim \frac{M^3_{\rm Pl}}{\Lambda^2}
\sim 10^{22}\,{\rm g} \, \left ( \frac{10^5\, {\rm GeV}}{\Lambda}
\right )^2,
\end{split}
\end{equation}
This PBH mass is not constrained by any observations, and PBHs in this mass range can account for all dark matter. 
Formation of PBHs occurs when the global charge from some critical radius, $R_*$ is collected in one Q-ball of the size $Q_{\rm max}$  required to form a single black hole, i.e.,
\begin{equation}
Q_{\max} = \frac{4\pi}{3}q_0 R_*^3,
\quad
R_* = \min \{H_*^{-1},m_\chi^{-1}\}	
.
\end{equation}
where $q_0$ is the charge density at fragmentation.  The second equation here reflects the possible Hubble-induced contributions to the scalar mass~\cite{Kawasaki:2011zi}. 
We can define the ratio
\begin{equation}
\Upsilon
\equiv
\frac{R_*}{H_f^{-1}}
=
\left(
\frac{Q_{\max}}{NQ_0}
\right)^{1/3}
.
\end{equation}
This ratio characterizes how much larger the Hubble radius must be, relative to that at the time of fragmentation, so that a black hole of size $M_{\rm BH}~=~\omega_c Q_{\max}$ can form after many iterations of mergers.

\begin{figure}[htb]
\includegraphics[width=0.95\linewidth]{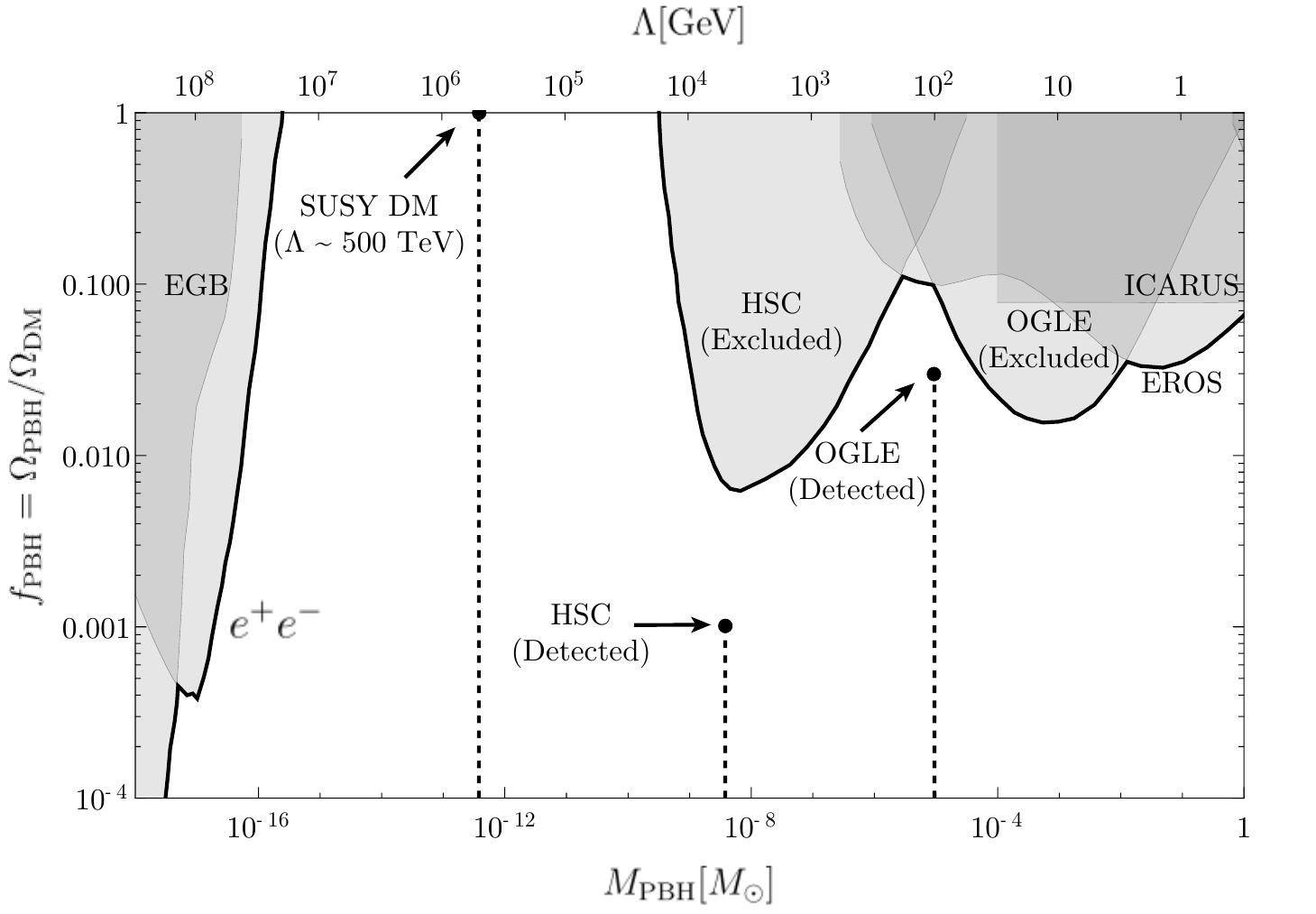}
\caption{The masses of black holes from supersymmetry with $\Lambda \sim 500\, {\rm TeV}$, as well as some heavier black holes that could be produced for smaller value of $\Lambda$. 
The constraints are from Refs.~\cite{Ali-Haimoud:2016mbv,Niikura:2017zjd,Niikura:2019kqi,Inoue:2017csr,Carr:2020xqk,Carr:2020gox,Lu:2020bmd, Dasgupta:2019cae, Laha:2019ssq,Sugiyama:2019dgt}. }
\label{fig:massfunction}
\end{figure}

Given that a single PBH will form inside a volume of $R_*$, the energy density of PBHs for scale factor $a$ is,
\begin{equation}
\rho_{\rm PBH}(a)
=
\frac{3}{4\pi}
\left(
\frac{M_{\rm BH}}{H_*^{-3}}	
\right)
\left(
\frac{a_*}{a}
\right)^3	
.
\end{equation}
where $a_*$ is the scale factor when $H^{-1} = R_*$.
The PBH abundance is usually expressed in terms of the fraction $f_{\rm PBH} \equiv \Omega_{\rm PBH}/\Omega_{\rm DM}$. Figure \ref{fig:massfunction} shows the current constraints on $f_{\rm PBH}$ across numerous mass ranges. The upper horizontal axis reflects the masses produced by various values of the energy scale $\Lambda$ as given by \eqref{eq:MPBHEq}. For example, assuming a ``flat-direction potential" with $\Lambda = 10^{5}\ \text{GeV}$ one finds $M_{\rm PBH}\sim 10^{-11}\ \text{M}_\odot$.

The abundance $f_{\rm PBH}$ is determined by the charge asymmetry $\eta_Q$ that remains in the condensate at the time of its fragmentation into Q-balls.  The flat direction may or may not be the one associated with the Affleck-Dine baryogenesis.  If there is no connection, then $\eta_Q$ is unrelated to the baryon asymmetry $\eta_B\sim 10^{-10}$.  

However, it is interesting to examine a more constrained possibility when the same flat direction is responsible for generation of baryon asymmetry and dark matter in the form of PBHs.  In this case, the condensate starts out with $\eta_{Q,i} \geq \eta_B$ at some high temperature $T_i>T_f$.  If the condensate decay rate into fermions is $\Gamma_Q$, then at the time of fragmentation the amount of global charge asymmetry left in the condensate is $\eta_Q = \eta_{Q,i}\exp(\Gamma_Q H_f^{-1})$.  One can verify that the right amount of dark matter is obtained in the case $\eta_Q \ll \eta_{Q,i} \approx \eta_B$, that is, when most of the baryonic charge from the condensate was transferred to the plasma fermions.   Because the scalar VEV is large, any fermion coupled to the flat direction is very massive, so the decay rate is exponentially suppressed, $\Gamma_Q \sim \exp(-1/\varepsilon)$, where $ \varepsilon = \omega/(g \langle \phi\rangle) \ll 1$, and $\omega$ is the energy per unit charge in the condensate~\cite{Dolgov:1989us,Pawl:2004cs}.  The PBH abundance from a decaying baryonic or leptonic flat direction is, therefore, 
\begin{eqnarray}
f_{\rm PBH}
&\simeq &
1.1\times 10^{9}\ \text{GeV}^{-1}
\cdot 
\frac{T_{\rm RH}  e^{-1/3\varepsilon}\Lambda^{1/3}}{Q_c^{1/12} H_f^{1/3}}\\[0.2cm]
&\simeq &
\left(\frac{T_{\rm RH}}{5\ {\rm GeV}}\right)
\left(\frac{e^{-1/2\varepsilon}}
{5.6\times 10^{-17}} \right)^{2/3}
\left(\frac{10^{32}}{Q_c} \right)^{1/12}\\[0.2cm] \nonumber
&\times & 
\left (\frac{\Lambda}{5\times 10^5\ \text{GeV}}\right )^{1/3}
\left (\frac{6\times 10^{-7}\ \text{GeV}}{H_f}\right)^{1/3}
\end{eqnarray}
where $T_{\rm RH}$ is the reheating temperature. For supersymmetry breaking scale $\Lambda \sim 500\, {\rm TeV}$ and $\varepsilon \sim 1.3\times 10^{-2}$, this gives the correct dark matter abundance.

One can also consider potentials of the same form, unrelated to supersymmetry, but realized in the dark sector.  Then the scale $\Lambda $ can take lower values, allowing for PBHs with the larger masses than those in Eq.~(\ref{eq:MPBHEq}).  Such PBHs can explain the microlensing events from HSC and OGLE~\cite{Niikura:2017zjd,Niikura:2019kqi,Sugiyama:2021xqg}.  
The mass functions for three sets of parameters are shown in Fig. \ref{fig:massfunction}. For all three cases illustrated, the coupling $y = 1$ GeV, the initial number of particles per horizon was chosen as $N = 100$ and the critical charge was set to $Q_c = 10^{32}$. The remaining parameters are specified in Table \ref{tab:paramtab}.

\begin{table}[htb]
\centering
\begin{tabular}{l|c|c|c|c}
\hline 
& $\Lambda$ {[}GeV{]} & $\varepsilon$  & $H_f$ {[}GeV{]} & $T_{\rm RH}\lesssim$ {[}GeV{]}   \\ \hline
Dark matter   & $5\times 10^5$ & $1.3\times 10^{-2}$ & $6\times 10^{-7}$ & 5\\
HSC  & $5\times 10^3$ & $1.2\times 10^{-2}$ & $3\times 10^{-10}$ & $2.3\times 10^{-2}$\\
HSC + OGLE & $1\times 10^2$ & $1.5\times 10^{-2}$ & $4\times 10^{-12}$ & $5.2\times 10^{-3}$\\
\hline
\end{tabular}
\caption{Model parameters}
\label{tab:paramtab}
\end{table}
These parameters also determine a bound on the reheating temperature. Should the $\chi$ field be a scalar field in MSSM, it may acquire a mass $\sim g T$ from thermal corrections. To avoid this, we assume the growth of PBHs occurs during reheating. However, even before the reheating is completed, the presence of radiation can affect the effective mass. We have evaluated this effect and require that the mass, with these corrections included, be still small enough for PBH formation.  This leads to the upper bound on the reheat temperature:
\begin{equation}
\begin{split}
T_{\rm RH}
&\lesssim
5\ {\rm GeV}\ 
\left(
\frac{\Lambda}{5\times 10^{5}\ {\rm GeV}}
\right)^{7/6}
\left(
\frac{H_f}{6\times 10^{-7}\ {\rm GeV}}
\right)^{-1/6}\\[0.25cm]
&\times
\left(
\frac{Q_c}{10^{32}\ {\rm GeV}}
\right)^{1/24}
\left(
\frac{e^{-1/2\varepsilon}}{5.6\times 10^{-17}}
\right)^{1/3}
.
\end{split}
\end{equation}
The maximum allowed reheating temperature is also specified in Table~\ref{tab:paramtab} alongside the corresponding parameters.

\section{Conclusion}

Primordial black holes present a natural and generic candidate for dark matter in supersymmetry. Supersymmetric generalizations of the standard model predict a large number of scalar degrees of freedom with a vanishing potential in the limit of exact supersymmetry.  Supersymmetry breaking lifts the flat directions in the potential by ``soft" terms (no quartic couplings), so that these directions remain relatively shallow and allow for a large values of the corresponding fields at the end of inflation.  The scalar condensate formed along these flat directions does not evolve homogeneously, but fragments into Q-balls, which can become the building blocks of PBHs.  While formation of PBHs from Q-balls is possible with gravitational interactions alone, it becomes much more likely when the Q-balls have additional attractive interactions mediated by scalar fields.  This class of scenarios, described in this {\em letter}, makes PBHs a likely and generic candidate for dark matter in theories with supersymmetry.  The dark-matter PBH mass is related to supersymmetry breaking scale as in Eq.~\eqref{eq:MPBHEq}, placing it in the open window for PBH dark matter.

\acknowledgments

This work  was  supported  by the U.S. Department of Energy (DOE) Grant No.  DE-SC0009937.    A.K.  was  also supported  by  the World Premier International Research Center Initiative (WPI),  MEXT,  Japan and by Japan Society for the Promotion of Science (JSPS) KAKENHI grant No.
JP20H05853, and by the UC Southern California Hub with funding from the UC National Laboratories division of the University of California Office of the President. 
A.K. thanks the Aspen Center for Physics, which  is  supported  by  National  Science  Foundation grant  PHY-1607611.


\bibliographystyle{JHEP}
\bibliography{bibliography}

\end{document}